\documentclass[12pt]{article}
\usepackage[USletter]{vmargin}
\usepackage[T1]{fontenc}
\usepackage[]{times}
\usepackage{color}
\usepackage{psfig,fancybox,fancyhdr,epsf}
\usepackage[]{fleqn}
\usepackage[dvips]{graphics}
\usepackage[english]{babel}
\begin{document}

\title{Stochastic measures and modular evolution in non-equilibrium thermodynamics}
\author{Enrique Hernández-Lemus \thanks{Corresponding author: ehernandez@inmegen.gob.mx}, Jesús K. Estrada-Gil\\ {\small Computational Genomics Department, Instituto Nacional de Medicina Genómica}\\{\small Periférico Sur No. 4124, Torre Zafiro 2, Piso 6, Álvaro Obregón 01900, México, D.F., México.}}
\date{}
\maketitle

\begin{abstract}

We present an application of the theory of stochastic processes to model and categorize non-equilibrium physical phenomena. The concepts of uniformly continuous probability measures and modular evolution lead to a systematic hierarchical structure for (physical) correlation functions and non-equilibrium thermodynamical potentials.\\

It is proposed that macroscopic evolution equations (such as dynamic correlation functions) may be obtained from a non-equilibrium thermodynamical description, by
using the fact that extended thermodynamical potentials belongs to
a certain class of statistical systems whose probability
distribution functions are defined by a {\it stationary measure};
although a measure which is, in general, {\sl different} from the equilibrium Gibbs
measure. These probability measures obey a certain hierarchy on its stochastic evolution towards the most probable (stationary) measure. This in turns defines a convergence sequence. We propose a formalism which considers the mesoscopic stage (typical of non-local dissipative processes such as the ones described by extended irreversible thermodynamics) as being governed by stochastic dynamics due to the effect of non-equilibrium fluctuations. Some applications of the formalism are described.

\end{abstract}

\section*{Scope}

The paper is outlined as follows: Section 1 is a brief introduction to the problem of applying measure theoretical tools to the study of many-particle physical systems, also some recent developments in the field are mentioned. We sketch how the probability measures approach has been applied to equilibrium systems (states). In section 2 we present an extension of such method for the case of non-equilibrium systems (processes) by means of the Choquet-Meyer Theorem on continuous measures. In this section two propositions (2.1 and 2.3) are made in terms of non-equilibrium stochastic measures as how to deal with systems out of local thermodynamic equilibrium within a modeling approach. The formalism dealing with Linear Response with Memory, (physical) Time Correlation Functions. Section 3 includes some results of the successful application of these methods on the aforementioned problems. Section 4 presents some brief concluding remarks.

\section{States and processes in Statistical Physics}

One of the great unfinished tasks of non-equilibrium physics,
whether {\sl statistical} or {\sl phenomenological} is that of
finding a general, model-independent way of describing
out of equilibrium states, similar to the Gibbs-Boltzmann formalism
for the statistical-kinetic description that leads to a complete
thermodynamical definition of systems in equilibrium. The Gibbs-Boltzmann formalism also applies to systems {\sl close} to equilibrium by means of the
celebrated {\it local equilibrium hypothesis}. This hypothesis lies at the heart of both linear response theory and linear (classic) irreversible thermodynamics.\\

The main problem in finding such a general theory of
non-equilibrium states lies in the fact that, due to their
intrinsic dissipative nature, such systems do not have a {\sl
unique} definition of entropy for each state \cite{meixent, eulib,
meixreol}, because of the non-volume preserving property of its
phase space \cite{otancaos, dorfman, gaspstat}. Since phase space
volume is not preserved it is, in general impossible to define a
unique global measure \cite{ruelam}, derive from this measure a
probability distribution function and characterize from it a
non-equilibrium system.\\

Nevertheless some work in that direction has been proved useful in recent years. For example, by extending the works of Sinai \cite{sinairu} and
later of Ruelle and Bowen \cite{ruelam, ruelbow} Gallavotti and
Cohen \cite{galaco, galaco2} wrote down an
expression for a non-trivial measure characteristic of a
stationary non-equilibrium state obeying this so-called SRB
statistics. This measure plus the so-called \textit{chaotic
hypothesis} make possible to formulate, at least at a formal stage
a proposal for a general theory of non-equilibrium phenomena. In this way probability measures give a new insight to our mathematical and physical understanding of non-equilibrium. \\

\subsection{Physical description and measurability}

Let us consider a description of a system in terms of its \emph{states}. For continuous systems the configurations are countable subsets $X$ of \textbf{R}$^{\nu}$ such that $X \cap \Lambda$ is finite for every bounded $\Lambda \subset $ \textbf{R}$^{\nu}$. A state should be defined as a probability measure on the set of all such $X$. We must give to each open set $\Lambda \subset $ \textbf{R}$^{\nu}$ the probability of finding exactly $n$ particles in $\Lambda$ and also the probability distribution of their positions. For every bounded open set $\Lambda$ and for every integer $n \geq 0$,  let $\mu_{\Lambda}^{n} \geq 0$ be a measure on $\Lambda^{n} \subset $ \textbf{R}$^{n \nu}$. We shall say that the $\mu_{\Lambda}^{n}$ form a \emph{system of density distributions} (SDD) if they satisfy the following conditions \cite{ruellerig}:

\begin{equation}\label{measero} \mu_{\emptyset}^{0} (\textrm{\textbf{R}}^{0}) = 1 \end{equation}

for every $\Lambda \subset \Lambda'$

\begin{equation}\label{measdens} \mu_{\Lambda}^{n}(x_1, x_2, \dots, x_n) = \sum_{p=0}^{\infty} \frac{(n+p)!}{n!p!} \int_{\Lambda'\setminus \Lambda} dx_{n+1} \dots \int_{\Lambda'\setminus \Lambda} dx_{n+p}\mu_{\Lambda'}^{n+p} (x_1,x_2, \dots, x_{n+p}) \end{equation}

The above conditions imply the normalization condition

\begin{equation}\label{normaliza} \sum_{n} \int dx_1 \dots dx_n \mu_{\Lambda}^{n} (x_1,x_2, \dots, x_n) = 1 \end{equation}

The (statistical) correlation functions\footnote{These statistical correlation functions are \textit{different} from the physical correlation functions that we will talk about later. The former are the moments of the SDD or Ensemble (as the physicist call it), correlating partitions of the probability measurable space; the latter are correlations between physical quantities which evolve in time. Obviously both kinds of correlation functions are related but their links are non-trivial, and in fact make up for a lot of the actual \emph{battlefield} of the statistical physicist.} corresponding to the density distribution $\mu_{\Lambda}^{n}$ are defined by:

 \begin{equation}\label{dens} \rho(x_1, x_2, \dots, x_n) = \sum_{p=0}^{\infty} \frac{(n+p)!}{p!} \int dx_{n+1} \dots dx_{n+p} \;\;\mu_{\Lambda}^{n+p} (x_1,x_2, \dots, x_{n+p}) \end{equation}

Here $x_1,x_2, \dots, x_{n} \; \in \; \Lambda$. Equation \ref{dens} could be formally inverted to obtain:

\begin{equation}\label{densmeas} \mu_{\Lambda}^{n}(x_1, x_2, \dots, x_n) = \frac{1}{n!} \sum_{p=0}^{\infty} \frac{(-1)^{p}}{p!} \int_{\Lambda^{p}} dx_{n+1} \dots dx_{n+p}\;\;\rho (x_1,x_2, \dots, x_{n+p}) \end{equation}

Since the series in the right-hand side of equation \ref{densmeas} may be non-convergent, there could be a case when the statistical correlation functions may not determine the state of the system. Nevertheless, it is usually asummed that the series converge.\\

It is possible to associate with every SDD $\mu_{\Lambda}^{n}$ a state on a $B^{*}$-algebra $\mathcal{A}$ that we will construct as follows:\\

For every bounded open set $\Lambda \subset$ \textbf{R}$^{\nu}$ let $\mathcal{K}_{\Lambda}^{\nu}$ be the space of real continuous functions on (\textbf{R}$^{\nu}$)$^{n}$ with support in $\Lambda^{n}$. Let us call $\mathcal{K}_{\Lambda}$ the space of sequences $(f^{n})_{n \geq 0}$ where $f^{n} \in \mathcal{K}_{\Lambda}^{\nu}$ and $f^{n}=0$ for large enough $n$. Let us now define $\mathcal{K} = \bigcup_{\Lambda} \mathcal{K}_{\Lambda}$.\\

We denote by $\mathcal{T}$ the topological sum $\sum_{n \geq 0}$ (\textbf{R}$^{\nu}$)$^{n}$  of disjoint copies of the spaces (\textbf{R}$^{\nu}$)$^{n}$. If $f =(f^{n})_{n \geq 0} \in \mathcal{K}$, a function $Sf$ on $\mathcal{T}$ is defined so that its restriction to (\textbf{R}$^{\nu}$)$^{n}$ is:

\begin{equation}\label{restric} Sf(x_1, x_2, \dots, x_n) = \sum_{p\geq 0} \sum_{i_1} ^{n} \dots \sum_{i_p}^{n} f^{p}(x_{i_1} \dots x_{i_p})  \end{equation}

Let us now consider functions on $\mathcal{T}$ of the form $\varphi(Sf_1, \dots Sf_q)$ where $f_1 \dots f_q \in \mathcal{K}$ and $\varphi$ is a bounded continuous complex function on \textbf{R}$^{q}$. With respect to the usual operations on functions and the complex conjugation operation (*), the $\varphi(Sf_1, \dots Sf_q)$ form a commutative algebra $\widetilde{\mathcal{A}}$ with an involution. The closure $\mathcal{A}$ of $\widetilde{\mathcal{A}}$ with respect to the uniform norm is an Abelian B$^{*}$-algebra \cite{ruellerig}.\\

Given an SDD $\mu_{\Lambda}^{n}$, a \emph{state} $\rho$ on $\mathcal{A}$ is defined as follows. For each $\varphi(Sf_1, \dots Sf_q)$, let $\Lambda$ be such that $f_1, \dots, f_q \in \mathcal{K}_{\Lambda}$. We write

\begin{equation}\label{rhorho} \rho(\varphi(Sf_1, \dots Sf_q))= \sum_{n \geq 0} \int dx_1 \dots dx_n \; \mu_{\Lambda}^{n} (x_1, \dots x_n) \times \varphi(Sf_1, \dots Sf_q) \end{equation}

$f_i = f_i (x_1,\dots x_n) \;\; \forall i$, this definition (equation \ref{rhorho}) is independent of the choice of the \textit{volume} $\Lambda$ and extends by continuity to a state on $\mathcal{A}$. If we call $\mathcal{F}$ the set of states defined as above (equation \ref{rhorho}) from an SDD; the mapping $\mu_{\Lambda}^{n} \to \rho$ is then one-to-one onto $\mathcal{F}$. It could be seen that the translations of \textbf{R}$^{\nu}$ are automorphisms of $\mathcal{A}$. The algebra here obtained $\mathcal{A}$ is not separable, hence a direct decomposition of invariant states into ergodic states is not possible. However, it can be shown that this could be attained if by an indirect treatment \cite{ruellemath}.\\

We are now on position of studying the time evolution of the just defined states (i.e. a physical \emph{process}) as a particular automorphism of the algebra $\mathcal{A}$.
The time evolution of the system occupying the finite region $\Lambda \in$ \textbf{R}$^{\nu}$, with respect to a given \textit{physical interaction}\footnote{Whose specific nature it will not affect the following consideration.} $\Phi$, is defined by a one-parameter group of automorphisms of $\mathcal{A}_{\Lambda}$ which associates to the \textit{observable} $A$ ($A(0)$) at time $0$, the following observable at time $t$ ($A(t)$):

\begin{equation}\label{modula} e^{itH_{\Phi}(\Lambda)} \; A \; e^{-itH_{\Phi}(\Lambda)} = \sum_{n=0}^{\infty} i^{n}\frac{t^{n}}{n!}[H_{\Phi}(\Lambda), A]^{n}
\end{equation}

here $[B, A]^{0} = A$; $[B, A]^{n} = [B,[B, A]^{n-1}]$.\\

It has been proved \cite{robinson} that in the so-called thermodynamic limit ($\Lambda \to \infty$), for a general class of physical interactions $\Phi$ with associated \textit{hamiltonians}\footnote{a Hamiltonian is a function which essentialy gives the energy of the system in terms of the positions of the particles and the nature of the interactions between them, in this case it is sufficient to know that is a bounded function of the $x_i$'s.} $H_{\Phi}$ the following limit exists:

\begin{equation}\label{limodula} \lim_{\Lambda \to \infty}\; e^{i \tau H_{\Phi}(\Lambda)} \; A \; e^{-i\tau H_{\Phi}(\Lambda)}= \Gamma_{\tau} \; A \end{equation}

Hereafter we will call $\Gamma_{\tau}$ the \emph{modular automorphism operator} and $\Gamma_{\tau} \; A$ will be called the $\tau$-modular evolution of $A$ or just the \emph{time evolution} of $A$.

\subsection{Equilibrium states, Gibbs measures and KMS condition}

Let us consider a quantum dynamical system described by a von Neumann algebra $\mathcal{A}$ (representing the system's physical observable quantities) with a one-parameter automorphism group (taking into account the time evolution of the observable quantities). In equilibrium statistical mechanics the entropy \footnote{In order to avoid the confusion between \emph{thermodynamic entropies} and probabilistic or informational entropies we will use term \emph{entropy} when referring to the former and \emph{entropic-measure} when referring to the latter.} $S$ of a given state $<\cdot>$ (labelled by an observable) is defined as:
\begin{equation}\label{quant} S = - \, Tr\; \hat \rho \, ln \hat \rho \end{equation}
where $\hat \rho = \rho / Tr \, \rho$ is the so-called normalized density matrix. We are to consider states lying on a constant observable energy  surface $E \,=\, <{\cal H}>$ with ${\cal H}$ the system's hamiltonian and the trace taken with respect to some measure $\Xi$ (an ergodic measure). For such systems, a global equilibrium state is characterizable as the state of maximal entropy within this set (of constant energy $E$) . As one knows the constant energy  and the normalization of the probability matrix $\rho$ ($Tr \hat \rho =1 $) constraints are introduced in the maximization procedure as Lagrange multipliers \cite{landau}. Doing so one gets the usual Gibbs states $\rho = e^{- \beta {\cal H}}$.\\

 It turns out that by analyzing the meaning of this entropy (making it coincident with the equilibrium thermodynamic entropy) one gets two main results. The parameter $\beta$ is proportional to the inverse temperature of the system and the ergodic measure $\Xi$ is identical to the Gibbs measure $\rho$ \cite{zubasm}. This equilibrium condition $\Xi \,= \, \rho$ (usually called Gibbs condition) has the restriction of being limited to apply to finite systems \cite{kubo}. \\

Now let us examine the quantum mechanical description of a system with an {\sl arbitrary} number of degrees of freedom under a constraint $\eta$ from the standpoint of a measure theory. The expectation value of an observable quantity $A$ in a state $\langle \cdot \rangle_{\eta}$ could be characterized by a density matrix $\rho_{\eta}$ as follows:

\begin{equation}  \langle A \rangle_{\eta} = \frac{Tr \, \rho_{\eta}A}{Tr \rho_{\eta}}\end{equation}

It is possible to introduce in connection with this measure a so-called {\sl modular Hamiltonian}  ${\cal H}^{\star}$ as:

\begin{equation} \rho_{\eta} = e^{-\beta^{\star} {\cal H}^{\star}} \end{equation}

Here $\beta^{\star}$ is a number introduced for later convenience (e.g. as a parameter of periodicity for the trace of density matrices). The modular evolution  \footnote{Of course it is possible to replace this modular evolution by the action of a Liouvillian super-operator or {\sl propagator}. See for example \cite{berne}, also \cite{mukamel}  For a modern introduction to modular operators see \cite{farkas}} of a property $A$ is given by an action-like similarity mapping \cite{zubasm} as:

\begin{equation} \Gamma_\tau (A) =  e^{\imath {\cal H}^{\star} \tau} A e^{- \imath {\cal H}^{\star} \tau} \end{equation}

The cyclical behavior of the trace (i.e. $Tr \rho_\eta (\tau) = Tr \rho_\eta ^{*}(\tau + \imath \beta^{\star})$) gives the following condition on the modular evolution between two states $N, M$:

\begin{equation}  \langle \Gamma_\tau (N) \, M \rangle_{\eta} = \langle M \, \Gamma_{\tau +\imath \beta^{\star}}(N) \rangle_{\eta} \end{equation}

This condition is called the {\sl KMS condition} \cite{kuboi}. This Kubo-Martin-Schwinger (KMS) condition (proposed by those authors as an auxiliary boundary condition for Green's functions) can be adopted as a \emph{simple characterizing property of equilibrium states} in the algebraic formulation (also called $C^{*}$-statistical mechanical formulation), which makes sense for an infinite system (in contrast to Gibbs Ansatz) and hence enables one to study an infinite system directly, by-passing the thermodynamic limit. This is a generalization of the theorem concerning Laplace transforms, which says that the Laplace transform of a function of energy E, which is zero for E < 0, is analytic as a function of the conjugate variable (here, time t) in the upper half-t-plane \footnote{For a proof of the theorem in the above context of modular evolution see reference \cite{robinson}.}. The KMS generalization of this says that, if the energy of a theory is positive then the \emph{equilibrium} state gives a two-point two-time correlation function that is analytic in time-difference in a strip in the upper-half plane of width $1/K^{\star}T$, where T is the temperature, and is periodic with imaginary period $1/K^{\star}T$, $K^{\star}$ is a measure-specific constant. The periodicity follows from the cyclicity of the trace. The justification of this KMS condition as the characterization of equilibrium states is that KMS states along with ground states and ceiling states (corresponding to $\pm 0$ temperature) are precisely those states which are stable against local perturbations.\\

 If the system is in a \emph{global} equilibrium state $\langle \cdot \rangle_\eta = <\cdot>$,  $\beta^{\star}$ is just the inverse temperature $\beta = 1/K_B T$ and the modular hamiltonian $\cal H^{\star}$ is the usual mechanical hamiltonian ${\cal H}$. In this case modular evolution generates time evolution. Stationarity and dynamical stability  are minimal requirements for a state to be called an equilibrium state; A passivity theorem  \cite{haag} can be rephrased by saying that it is precisely the KMS states which are distinguished by the second law of thermodynamics in Kelvin's formulation: there are no cyclic processes converting heat into mechanical work if the state of the systems obeys the KMS condition \cite{roos}.\\

Local equilibrium could also be defined through a KMS-type condition. We can think of local thermodynamic equilibrium as a state that cannot be distinguished from a global equilibrium state by infinitesimally localized \emph{measurements} \cite{hess}. The quantitative description of the term \textsl{infinitesimally localized measurements} could be given in terms of a one-parameter scaling procedure  \cite{hess} over the observable. We introduce a parametric diffeomorphism as a scaling procedure as follows.

\subsection*{Definition 1.3\\ \emph{Uniparametric state space scaling}}
\textit{A diffeomorphism of the state space that preserves the topological and ergodic character of such space and projects a set of points (called quasi-local points) into equilibrium points by means of a change of scale is called a Uniparametric State Space Scaling (USSS)}.\\

Let us consider an observable depending on n-points in real space $A=A(x_1, x_2, \dots, x_n)$. For the sake of simplicity we will consider a unique field $\phi$ spanned by the various values of the observable and depending on spacial localization as follows: $A=\phi(x_1) \phi(x_2) \dots \phi(x_n)$. A $\xi$-scaling diffeomorphism (USSS) $\Lambda_\xi$ is given as:

\begin{equation}\label{scaling} \Lambda_\xi A = \Sigma(\xi)^{n} \, \phi(\chi_\xi x_1)\dots \phi(\chi_\xi x_n)  \end{equation}

with $\Lambda_\xi A$ the value of $A$ after a scaling by a factor $\xi$, $\Sigma(\xi)$ a scaling function (postulated in order for the scaling to be \emph{well-defined}) and
\begin{equation} \label{esca} \chi_\xi (x_i) = x^{*} + \xi(x_i - x^{*}) \end{equation}

$x^{*}$ is the point to which quasi-local points \footnote{We call quasi-local to those points with a value of the property A sufficiently close to its equilibrium value $A^{*}$ so that after a finite scaling they will indeed take the value $A^{*}$. Hence quasi-local points are \textsl{good candidates} for representing local equilibrium states.} $x_i$ shrink after \textsl{localization}, $x^{*}$ is called an \emph{equilibrium point-cell} in the sense of Onsager. It is possible to see that $\chi_{\xi=0}(x_i) = x^{*} \; \forall i$  and that $\chi_{\xi=1}(x_i) = x_i$ so that if the system scales with $\xi \to 0$ we could talk about local equilibrium (i.e. all points shrink to an equilibrium one) and if $\xi \to 1$ we are in a highly delocalized stage and hence local equilibrium is not attained. Of course repeated application of the scaling $\Lambda_\xi$ would define a \textit{coarse graining operator} in the sense evoked by Ehrenfest and formalized by Zwanzig \cite{zwanzigop}.\\

It is worth noticing that this scaling procedure generates, for each application, a sequence of probability distributions $\rho_0,\rho_1, \dots \rho_n \dots$; here $\rho_0$ is the initial random distribution, generally taken as a uniform distribution, a fact called \emph{Stosszahlansatz} (by Boltzmann), molecular chaos hypothesis or Initial Random Phase Approximation. Since the USSS procedure is, by construction, convergent to the state of thermodynamic equilibrium, $\rho_{\infty}$ should be the equilibrium distribution function, i.e. the equilibrium Gibbs distribution,  $\rho_{\infty}=\rho$. A weak convergence condition is generally assumed:

\begin{equation}\label{convergence} \lim_{n \to \infty} \rho_n = \rho \end{equation}

Even if the state of equilibrium (global or local) is not attained the description of the system under a modular evolution could be done. In this case the measure will depend on the {\sl space-time localization} described by the scaling $\Lambda_\xi$. For a variety of conditions a systematization for the scaling effect could be done in terms of a time-dependent distribution function $\Omega_{\tau}(\xi)$ for the values of $\xi$. The \textit{time} $\tau$ can be considered as the indicator for the direction of convergence. In some cases the distribution function is given in the form of a stationary absolutely continuous uniform measure (or a family of Cauchy convergent continuous measures) and so the weak convergence argument -equation \ref{convergence}- applies. \cite{galaruel,absmeasure}.\\

Several physically relevant problems could be better understood in terms of uniformly continuous measures of parameter state spaces. The problem of extending the notion of Gibbs states to quasi-local interactions \cite{leny}, coupled quantum systems \cite{jaksi},  the hierarchical structure of physical descriptions in terms of the coarse-graining \cite{guido} exemplifies such problems. The recent arise of the so-called {\sl non-extensive  statistical mechanics} \cite{tsallis} and the somehow dubious {\sl q-based entropies} \cite{physallis} are an indirect consequence of a problem that has been called {\emph non-canonical averaging} \cite{michelle}. This problem has been related with the issue of {\it aging} in physical systems in the form of anomalous phenomena such as the ones surrounding the glass transition, a problem that could also be treated as the effect of a non-local in time measure. Quasi-local interactions, coarse-graining-scale effects and {\sl non-canonicity} are all in a class of phenomena for which the description in terms of absolutely continuous measures (phase space densities) results enlightening.\\

As we already noticed a central issue in non-equilibrium physics is the description of irreversible processes in a formalism akin to the Gibbs-Boltzmann Ansatz. Since we have seen that a measure-theoretical description carries on the double duty of characterizing equilibrium (and also local equilibrium) states through a KMS condition, and also providing a tool for calculation of the evolution of equilibrium quantities (i.e. modular evolution via an equilibrium distribution called the Gibbs measure $\rho$) it is natural to think of a possible extension of this procedure for the characterization of non-equilibrium processes. Nevertheless, even if this goal has been pursued by some of the most brilliant minds in thermal physics: Boltzmann, Landau, Onsager, Zwanzig, Green and many others; the problem has been proved to be very difficult to tackle. The reason is that since we have to base our measure theoretical description on a non-stationary hamiltonian (a strongly time-dependent quantity), the explicit form of the associated measure is given in terms of the solution (impossible in practice) of the complete many-body problem. Several approaches have been attempted, ranging from projection operator techniques \cite{zwanzigop,zwanzig65} to memory functions \cite{yipboon}, variational principles \cite{zubasm}, etc.\\

In the next section we will introduce an alternative based on a kinetic-theory-founded \cite{eujcpd95} extended irreversible thermodynamical formalism \cite{eulib,eumat,eupr95} taking into account the fact that in the typical time and length scales of extended non-equilibrium phenomena (the so-called mesoscopic stage) the fluctuations of the macro-variables will play a very important role.

\section{Non-equilibrium measures}

By application of the so called {\sl modified moment method} for the kinetic theory extraction of macroscopic non-equilibrium information, it has been proved \cite{eujcp86} that a {\sl probabilistic} measure (a "quasi-gibbsian" measure) exists for an arbitrary  pair of (non-equilibrium) steady states given in terms of the entropy production (more precisely the uncompensated heat production) at the steady states. This measure gives rise to an {\sl extended Gibbs relation} that generalizes the Gibbs relation for the entropy change usually employed in equilibrium thermodynamics \cite{eumat}. \\

The distribution function $f^{\small ne} = f_0 + \, f^{\small ne}_1 + \, f^{\small ne}_2+ \dots$, if examined under the so-called functional hypothesis \cite{eujcpd95, eujcpm95} gives rise to an irreversible thermodynamical formalism that, in spite of being of a nonlocal and nonlinear ({\sl far from equilibrium}) nature obeys a form of \textsl{canonical} thermal averaging; this fact  will be very useful later in this study. Here $f_0$ is the equilibrium distribution function and $f^{\small ne}_i$, $i = 1,2, \dots$ are non-equilibrium contributions to the distribution function also called {\it higher order moments} of the distribution function.\\

The specific form of the measure $\mu$ depends on the solution of the kinetic theoretical description of the system under study. Once again, except for a few ideal systems like the classical and quantum diluted gas there is no closed solution for the kinetic equations. However, since the measure depends on the kinetic solution in terms of a distribution function $\mu = \mu \,(f_0, \, f^{\small ne}_1, \, f^{\small ne}_2, \dots)$ (even without having the explicit solution for it), it is possible to see that the passage from the so-called {\sl chaotic stage} (given by Boltzmann's stosszahlansatz) to the fully-developed stationary distribution shall induce a {\it dynamic convergence sequence} in the measure. Given these facts the main contribution of this work could be summarized in two principles:

\subsection*{Proposition 2.1}

\emph{The measure $\mu_\tau$ associated with the state space modular evolution $\Gamma_\tau (\vec X)$ for a system out of (local or global) thermodynamic equilibrium  and described by variables $\vec X (t)$ is given by a  sequence of time dependent distribution functions. This means that we are able to recognize the measure as the stationary solution of a stochastic process $\Theta_\tau (t)$.}\\

More explicitly we can state that:

\begin{equation} \label{estocass} \lim_{t \uparrow \tau} \Theta_\tau (t) = \mu_\tau \end{equation}

\begin{equation} \label{propa} \Gamma_\tau A = \int_{\Omega_\tau} \Theta_\tau (t) \,  F \left(A(t), \tau \right)\, dt = \int_{\tau} \mu_\tau F^{\tau}\left(A(t)\right) \, dt    \end{equation}

$F$ is called the kernel or modulus of the given modular evolution $\Gamma$ and is a still-undefined \footnote{System-dependent in the Physicists saying} continuous monotonic function of the state space fields $\vec X$. The dynamic variable $A$ is an arbitrary function of the state space fields $\vec X$, $A(t)\equiv A \left(\vec X(t)\right)$, $\Theta_\tau$ is a time-continuous stochastic process. The sequence induced by equation \ref{propa} converges according to the distribution $\Omega_\tau$ already mentioned in connection with the scaling procedure in the previous section. Although the explicit form for the distribution function for the scaling parameter $\Omega_\tau$ is \textit{a priori unknown} we will see that is homeomorphic to the convergence sequence for the kinetic distribution function $f^{ne}$.

\subsection*{Theorem 2.2}
\emph{The stationary state of the Stochastic process $\Theta_\tau$ corresponds to the non-equilibrium probability measure $\mu$ at least in the sense of means (i.e. almost-surely) according to equation (\ref{estocass})}.

\subsection*{Proof}
\emph{Let} $\Omega_{\tau}(\xi)$ \emph{be a distribution of} $\tau$\emph{-time dependent scaling factors in the sense of USSS. Let us assume that the distribution is stationary and has a compact support (i.e. it is \textit{dense} in state space). By construction of the USSS, the stationary distribution corresponds to an equilibrium (local or global) distribution of state space points. Now let }$\mu = \mu \,(f_0, \, f^{\small ne}_1, \, f^{\small ne}_2, \dots)$ \emph{be an} $\epsilon$-\emph{dependent or kinetic distribution, where} $\epsilon$ \emph{is called the uniformity parameter or Knudsen's parameter. By definition} $\mu$ \emph{is stationary and has compact support in phase space. Since the stationary distribution} $\mu\,(f_0)$ \emph{is an equilibrium (local or global) distribution there exists a transformation from one equilibrium distribution to the other (i.e. in the thermodynamic limit both distributions are equivalent).}\\

\emph{This condition suffices, to take into account the fact that as }$\xi$\emph{ goes to zero the kinetic distribution} $f^{ne}$ \emph{should converge to the equilibrium distribution }$f_0$. \emph{This is so, since the condition of equilibrium (in the sense of spatial homogeneity of the thermodynamic potentials) gets attained (see equation \ref{esca}), notwithstanding the difference in the speed of convergence of both distributions}.\\

In some sense, \emph{Theorem 2.2} is a kind of informal version of the Monotone-Convergence Theorem \cite{wiliams}. The second proposition, namely equation (\ref{propa}) follows \emph{Theorem 2.2} plus the assumption of modular evolution. Let us see this in terms of measures:\\

Let $F$ be a locally convex topological vector space and $K$ a convex, compact subset of $F$. The dual $\mathcal{M}$ of the closure of $K$ consists on the measures on K. Denoting by $\mathcal{M}_+$ the convex cone of positive measures and by $\mathcal{M}_1$ the set of positive measures of norm 1 (i.e., the probability measures). If $\mu \in \mathcal{M}_1$ there exists  $\rho \in K$, such that:

\begin{equation}\label{intemu} f(\rho) = \int f(\sigma) d\mu(\sigma) \end{equation}

$\rho$ is usually called the resultant of $\mu$ \cite{burba}. If $\mu \in \mathcal{M}_1$ has a resultant $\rho$ then $\mu$ can be approximated weakly by measures $\mu' \in \mathcal{M}_1$ with resultant $\rho$ and finite support \footnote{A formal proof of this \emph{Choquet-Meyer Theorem} (too lengthy to be included here) is given in reference \cite{choquet}}\\

If we take this result in terms of equation \ref{estocass}, equation \ref{propa} follows directly. In the second expression at the right hand side of equation \ref{propa} has been used the so-called \emph{tower property of conditional probabilities} \cite{wiliams}. \\

We have still retained the notation of discrete evolution in order to show up the analogy with usual (i.e. finite volume) Gibbs measures, however we must note that $\mu_\tau \, F^{\tau}$ represents the operation of a time continuous stochastic process $\Theta_\tau (t) $ over a field through the action of a $\tau$-continuous semigroup $F$ \footnote{Since  $F^{\tau}$ represents the time evolution of a \emph{mesoscopic}  and hence possibly dissipative system the inverse operation may not satisfy existence and/or unicity so we will refer to the effect of operating $F$ as the action of a semigroup.}.\\

Once we have defined this measure we are able to make explicit assumptions to what the time evolution of a macro-variable will be in terms of correlation functions under the context of non-local irreversible thermodynamics.

\subsection*{Proposition 2.3}

\emph{The time evolution of a (non-local) irreversible thermodynamic field Z given in terms of a modular evolution operator $\Gamma_\tau$ generates a form for the 2-time correlation function given by the $\tau$-continuous stochastic map:}

\begin{equation}\label{evol} \langle Z(t), \, Z(t') \rangle_{EIT} =  \int_{-\infty} ^{t} Z(t) \, \mu_{(t-t')} \, Z(t') \, dt'\end{equation}

\emph{Similar expressions could be write down for higher order  and crossed correlations.}\\

As it could have been already noticed \emph{Propositions 2.1} and \emph{2.3}, are an extension of linear response theory for the (non-linear) case of a time dependent measure. At the moment and having recognized the impossibility to derive this measure from microscopic models we have assumed it as an stochastic variable. Of course if this stochastic measure converges (at least in the sense of means) -see equation \ref{convergence}- to the Gibbs measure $\rho$ we recover the usual linear response theory.

\subsection{Linear response with memory}

Non-equilibrium thermodynamics often addresses the problem of transport through material media. Hyperbolic transport equations (such as the MCV \cite{cattaneo,vernotte} equations and the telegrapher's equation \cite{oliva}) taking into account the lag on the response due to the finite velocity of perturbations have been derived from several points of view. Ranging from phenomenological arguments to purely microscopic transport \cite{oliva} and also probabilistic methods such as the persistent random walk \cite{gold}. On the other hand linear response theory represented a powerful theoretical tool to cope with transport phenomena from a dynamic and thermodynamic standpoint. In the last years linear response functions incorporating \textit{memory}  (i.e. the effect of the lag in the response to an applied field) appeared in such problems as delayed transport in electronic devices, molecular hydrodynamics and rheology of structured fluids, and also solid state phenomena such as the dynamics of Abrikosov vortices.\\

Let us first consider two related dynamic variables say the magnetic field $\vec {\cal H}(t)$, and the magnetization $\vec{\cal M}$. At constant temperature in the low-field limit there is a linear relation between the magnetic field and the magnetization:

\begin{equation}\label{reslin} \vec{\cal M} = \chi_T^{0} \; \vec {\cal H}(t) \end{equation}

Here $\chi_T^{0}$ is a tensor response coefficient called the susceptibility tensor. However, it is known that it takes some time $\tau_{M}$ for a material media to achieve magnetization under exposure to a magnetic field. In other words, there is a lag in the response to the field. This effect is due to the fact that magnetization is a non-equilibrium process. If we look at the effect of mesoscopic fluctuations of the magnetic field due to non-equilibrium evolution in the sense already evoked we will see a possible explanation. Since the magnetic field is in disequilibrium we can assume it undergoes a stochastic time evolution. It is possible to associate an automorphism $\Gamma_{\eta}$ on the magnetic field with this stochastic evolution:

\begin{equation}\label{automes} \vec {\cal H}(t)_{sto} = \Gamma_{\eta} \vec {\cal H}(t) \end{equation}

If we consider linear response between the magnetization $\vec {\cal M}$ and the field $\vec {\cal H}(t)_{sto}$ we get:

 \begin{equation}\label{reslin2} \vec{\cal M} = \chi_T^{0} \; \vec {\cal H}(t)_{sto} = \chi_T^{0} \, \Gamma_{\eta} \vec {\cal H}(t) \end{equation}

But since the modular automorphism is induced by a time-continuous measure $\mu_{\tau}$ we have:

\begin{equation}\label{reslinmes} \vec{\cal M} = \chi_T^{0} \; \int_{-\infty}^{t} \mu_{\tau} (\vec {\cal H}(t'))\, dt' \end{equation}

By proposing a form for the stochastic process associated with the measure $\mu_{\tau}$ of the automorphism $\Gamma_{\eta}$ we can \textsl{model} this non-equilibrium process. If we look for the behavior in a weak stochastic limit we are able to consider a first order stochastic process. Let us consider $\Gamma_{\eta}$ as a modular automorphism whose associated stochastic process is a simple exponential decay in time since this is a prototypic first order stochastic process \cite{kampen}. In this case equation \ref{reslinmes} reads:

\begin{equation}\label{reslin3} \vec{\cal M} = \chi_T^{0} \; \int_{-\infty}^{t} \lambda_{P} \; e^{(\frac{t-t'}{\tau_{P}})} \; \vec {\cal H}(t')\, dt' \end{equation}

here $\lambda_{P}$ is the amplitude of the associated distribution and $\tau_{P}$ is a characteristic time (akin to Poisson's time). If we define a non-equilibrium susceptibility $\chi_T^{ne}$ (up to first order) as $\chi_T^{ne} = \chi_T^{0} \lambda_{P}$, and recognize Poisson's time as the relaxation time associated with the non-equilibrium process of magnetization we get:

\begin{equation}\label{reslin4} \vec{\cal M} = \int_{-\infty}^{t} \chi_T^{ne} \; e^{(\frac{t-t'}{\tau_{M}})} \; \vec {\cal H}(t')\, dt' \end{equation}

Equation \ref{reslin4} is the usual expression of a linear response with exponential memory (also called fading memory). If we look at the differential representation of equation \ref{reslin4} we will find the hyperbolic MCV-type transport equation. It is worth noticing that even if we start with a linear response relation and the measure is induced by a first order stochastic process, the resulting equation possess a {\it non-markovian} character, i.e. memory in the response. If the time scale of the system is much more slower than its stochastic time ($t_{char} >> \tau_{P}$) then we can consider the limit $\tau_P \to 0$. In this case the exponential decay distribution is just a $\delta$-distribution and the usual linear response without memory is recovered. By means of the analysis of the associated measure (and its related stochastic process) one is enable to perform a \textsl{selection} of the effects that we will take into account in a very transparent way. This will be a very important issue when considering a collection of irreversible processes taking place in a non-equilibrium system, since in this case the presence of irreversible couplings (a question related to the relative relaxation times of the various processes) could be cope-with in a systematic way.

\subsection{Time correlation functions}

Time  correlation functions are usually considered to represent the mean
behavior of a large set of microscopic dynamic variables under certain
averaging assumptions (canonicity, number density conservation,
energy density conservation, etc.) usually taken into account by some
kind of Lagrange multiplier formalism (in the case of extended thermodynamics these multipliers are introduced, for example, by Liu's Method) \cite{eustatme,joucasleb,leboncas}. Nevertheless by applying a novel irreversible thermodynamics formalism \cite{eumolscat}
it has been stated  that they also represent Gibbsian ensemble averages of
a collection of macroscopic field variables, when the latter are
considered (as is the case in extended irreversible thermodynamics
\cite{eulib}) as coming from stochastic processes (in general
non-markovian noises) in a mesoscopic length-scale \cite{eurinv,eujcp86}.
This statement has only been proved formally for dilute gaseous
systems by means of quantum kinetic theory \cite{eulib,eurinv,eujcp86},
but there is strong evidence supporting its validity for several other
physically significant systems. We will use this statement as an ansatz.\\

In order for the field averages to be of a more general character, we will use the equivalence between 2-time correlation functions and {\sl propagators} \cite{zwanzigop, zwanzig65}, or time evolution operators acting on a dynamical variable. The {\sl dynamic variables} are the set of extended thermodynamical potentials over its gibbsian set. {\sl Time evolution operators} are obtained as inner products with {\sl dual vectors} of the aforementioned thermodynamical potentials. From the theory of stochastic processes \cite{kampen, martin} we obtain the statistical properties of such {\sl inner products} by introducing a probability measure $\mu$ (or {\sl weighting function}) in the last stages of study of the problem.\\

The 2-time auto-correlation function for a dynamical variable $A$ will then be given as (see eq. \ref{evol}):

\begin{equation}\label{2corre} \langle A(t) A(t') \rangle_{t'}  = \int_\tau A(t) \mu_\tau A(t') \, dt' \end{equation}

As we already stated we are in a position to see the measure $\mu_\tau$ as a dual complement of $A$, i.e. $\mu_\tau A = A^{\dag}$ with $(A(t), A(t)^{\dag}) = \| A (t)\|^{2}$    \cite{stroock}. We could also look in a very similar way to 2-time crossed-correlation functions and also to higher order correlations. Let us look at an example coming from extended irreversible thermodynamics \cite{artitesis}.\\

According with the formalism of Extended Irreversible Thermodynamics (EIT) \cite{joucasleb} the time evolution of the (entropy-like) {\sl compensation function} $\Psi$ is given by:

\begin{equation}\label{egibbs}T d_t \Psi = d_t U +  P d_tV - \sum_i {\Upsilon_i d_tC_i} + \sum_j{X_j \odot d_t\Phi_j}\end{equation}
 We see that equation (\ref{egibbs}) is nothing but the formal extension of the celebrated Gibbs equation of equilibrium thermodynamics for the case of a multi-component non-equilibrium system. This extended Gibbs relation is brought about in the theory by imposition of some {\sl consistency conditions} on the non-equilibrium part of the distribution function \cite{eumolscat,eurinv, eujcpd95}. The quantities appearing therein are the standard ones, $T$ is the local temperature, $P$ and $V$ the pressure and volume, $\Upsilon_i$ is the chemical potential for the species "$i$", etc. $X_{j}$ and $\Phi_{j}$ are extended thermodynamical fluxes and forces.\\

We will look up for the effect that a non-equilibrium process, say mass flux will have on the thermodynamic description of the system.  In the case of a binary fluid mixture we will take our set of relevant variables ${\cal{G}} = \cal{S} \, \bigcup \, \cal{F}$ to consist in the temperature $T(\vec r, t)$ and concentration of one of the species $C_2(\vec r, t)$ fields as the slow varying (classical) parameters set $\cal{S}$ and the mass flux of the same species $\vec J_2 (\vec r, t)$ as a fast variable on the extended set $\cal{F}$.
For the fast dynamic variables (such as the mass flux $\vec J_2$ and its conjugated {\it thermodynamic force} $\vec X$) characteristic times are much smaller than for the so-called conserved fields (mass, energy and momentum densities, etc.) so, the effect of fluctuations in this variables will be greater. We will take this fact into account by associating to this fields a memory in the response akin to the memory described in the last section. \\

We propose linear  constitutive equations with exponential memory kernel (see section 2.1) for the following reasons: a) The associated transport equations are hyperbolic (of the Maxwell-Cattaneo-Vernotte type) \cite{cattaneo, vernotte} so causality is taken into account, b) These hyperbolic transport equations are compatible with the postulates of EIT  \cite{joucasleb,leboncas, muschik, muschikasp, muschikrec}, c) Similar equations can be derived for coupled non-markovian stochastic processes \cite{weizs, intepro} and since stochasticity has been associated with the major role of fluctuations in the mesoscopic description level of EIT the outcome will improve our understanding of phenomena occurring in such mesoscopic stages, d) The stochastic process associated with this set of two coupled constitutive equations called a \textsl{semimartingale} has been extensively studied and its known to accept absolutely continuous measures a fact that will become very useful \cite{absmeasure,wiliams,martinga}.\\

The constitutive equations are therefore chosen to be,
\begin{equation} \label{jota1} \vec J_2 (\vec r,t) =  \int_{-\infty} ^{t} \lambda_1 \, \vec u \, e ^{\frac {(t'-t)}{\tau_1}}
\Upsilon ^\dagger (\vec r,t') dt'\end{equation}

\begin{equation}\label{equis1} \vec X ^\dagger (\vec r,t) = \int_{-\infty} ^{t} \lambda_2 \, e ^{{(t''-t)}\over{\tau_2}}
\vec J_2 (\vec r,t'') dt'' \end{equation}

The $\lambda_i$'s are time-independent, but possibly anisotropic amplitudes, $\vec u$ is a unit vector in the direction of mass flow and $\tau_i$'s are the associated relaxation times considered path-independent scalars. Since we have a linear (thought \textit{non-markovian}) relation between thermodynamic fluxes and forces some features of the Onsager-Casimir formalism  will still hold. The resulting thermodynamic potentials and its derivatives are assumed to belong to the quasi-gibbsian set of time dependent macroscopic functions mentioned in reference \cite{eujcp86}, over which we will perform a thermodynamic average (i.e. a phase field average as measured by a quasi-gibbsian measure characteristic of the non-equilibrium thermodynamic state space) in order to obtain the 2-time dependent correlation function for, say the concentration-concentration correlation function $\langle C_1(\vec r,t),C_1(\vec r,t')\rangle$. If the thermodynamic average is performed under isotropic canonical conditions \cite{zubasm, zubarusm, hurley} the resulting correlation function is given by: \\

\begin{equation} \label{nneitvh}
\langle C_1(\vec r,t) , C_1(\vec r,t') \rangle = \int_{\Phi} C_1(\vec r,t)\; \mu_{(t- t')}C_1(\vec r,t') dt' \end{equation}

For the case being treated it was showed \cite{artitesis} that equation \ref{nneitvh} is given as follows:

\begin{equation} \label{eitvhtes}
\langle C_1(\vec r,t) , C_1(\vec r,t') \rangle = \int_{\Phi} \Xi(\vec r) \; e ^{-n (t+t')} \; W(t - t')dt' \end{equation}

with
\begin{equation}
\Xi (r) = AB + B ^2
\end{equation}
\begin{equation} A = \int_{o} ^{T}{\frac{C_p(T)}{T}} dT \, {\frac{1}{H(\vec r)}} \end{equation}

\begin{equation}
B=[{{\lambda_1 \lambda_{2}  \tau_1 H(\vec r) T(\vec r)}\over{n(1-n \tau_1) ^2}}]
[\lambda_1 \tau_1 ^2 - \tau_2 ({{\lambda_1 \tau_1}\over{(1-n \tau_1)}}+{{ \lambda_1 (1-n \tau_1) -1}\over{(1-n \tau_2)}}) ]
\end{equation}

$W(t-t')$ is a stochastic process representative of the measure $\mu_{(t-t')}$; hence the presence of non-equilibrium correlations is taken into account by means of this stochastic evolution of the measure \footnote{In the case studied then (reference \cite{artitesis}) the convergence sequence was Cauchy, with stationary {\sl Ornstein-Uhlenbeck} character for linear non-markovian statistics except in the case where stochastic resonance was present.
}. Equation (\ref{eitvhtes}) is the non-regular part of the composition field time correlation function (i.e. the value of the correlation function near the critical point after scaling of the thermal and concentration fields). $C_p$ is the heat capacity, $T(\vec r)$ and $H(\vec r)$ are known continuous functions for the amplitudes of the temperature field and the chemical potential field. \\

For a broad family of stochastic measures equation (\ref{eitvhtes}) asymptotically converges to a limit given by \cite{ryzhik}:

\begin{equation} \label{convh}
\langle C_1(\vec r,t) , C_1(\vec r,t') \rangle =  \kappa \; \Xi(\vec r) \; e ^{-2n t}  \end{equation}

where $\kappa$ is a constant depending on the explicit stochastic measure under consideration. For an \emph{unitary Ornstein-Uhlenbeck} measure ($W(t-t') = e^{|t-t'|}$) convergence implies $\kappa = {\frac {1}{1-n}}$. As we shall see this distribution resulted very appropriate to model this kind of non-equilibrium critical system \cite{artitesis}. In the case of an \emph{unitary Gaussian measure} ($W(t-t') = e^{(t-t')^{2}}$) for example, an asymptotic solution could not be given in terms of an exponential decay, a typical feature of the critical decay of fluctuations. In this Gaussian case the solution consist of two contributions: an exponential decay plus an error function type mode. This difference eliminates the short-time plateau (known as critical slowing down of fluctuations) present in correlation functions obtained under colored noise measures. \\

Equation \ref{convh} is a form of the van Hove expression for the density-density correlation function, also called a dynamic structure factor.\\

One of the advantages of this formalism with respect to others, such as
mode-coupling theory  is that we can test different kinds of stochastic relaxational couplings (delta-correlated, quasi-markovian, gaussian,  lorentzian, Ornstein-Uhlenbeck type, non-markovian, simultaneous non-linear multi-modal coupling, etc.) just by changing the weighting functions $W(t-t')$ (that is the family of probability measures $\mu_\tau $), within a unifying thermodynamic scheme.\\

\subsection{Hysteresis in non-equilibrium systems}

It is possible to describe how the phenomena of hysteresis could be well understood in the context of an extended irreversible thermodynamic formalism by pointing out that hysteretic
phenomena is a consequence of the existence of dissipative internal processes in the system as was showed elsewhere \cite{histe}. The different contributions to the uncompensated heat production can be explicitly incorporated in the description by characterizing each one by its relaxation time. In order to do so it is important to take into account the effect of dynamical coupling. One way of taking this effects into account is by examining the coupling of the associated stochastic measures, then looking for processes whose measures have stochastic times of the same order, since in this case a dynamic coupling is possible. By {\sl enslaving} those stochastic processes to the faster one it is possible to develop a hierarchy of measures. In this case the analysis of relaxation times is also physically clearer than by using the standard {\sl hysteresis operators} \cite{prei,maye}

\section{Some successful examples of the results of the application of the formalism}

Since we introduced the non-equilibrium stochastic measure $\mu_\tau$ in Propositions 1 and 2 as a tool for {\sl modeling} systems out of local or global equilibrium its \emph{physical} validity will be ultimately conditioned by the results given by its use against experimental data. We will give here some brief results of some cases where its application has led to satisfactory results. Let us first consider the extended thermodynamical description of criticality as given in ref.\cite{artitesis}. If we compare the results for the temperature dependence of the diffusion as obtained from the calculated 2-time concentration autocorrelation function in the neighborhood of the critical consolute point, as given by equation \ref{eitvhtes} under a suitable measure with the experimental light scattering spectra as reported and by looking at the results there, it is clear that the stochastic measure was a reliable tool for the description of a highly complex non-equilibrium property. \\

For the case of non-equilibrium couplings and its relation to the observed phenomena of hysteresis (cf. section 3.3) it has been possible to show \cite{histe} that if the relaxation times of the phenomena are compared to the {\sl stochastic time} given by the associated measure, then a criteria is given as to when will we experimentally observe hysteretic phenomena.\\
 
 In brief; by applying simple and reasonable modeling approximations within the formalism outlined in the preceeding sections, we were able to reconstruct the hyperbolic transport MCV equations (cf eq. \ref{reslin4}) of linear response theory with memory. Also a well known result of condensed matter theory and phase transitions was obtained, namely the Van Hove equation for the density autocorrelation function (eq.\ref{convh}). Finally some previous results on the phenomena of hysteresis in terms of relaxation times resulted firmly grounded within the framework presented in this paper.

\section{Concluding remarks}

In equilibrium statistical mechanics a measure theoretical representation (Gibbs formalism) has provided a great insight and also powerful tools for the characterization of equilibrium states (via Gibbs or KMS conditions) and the calculation of equilibrium quantities (via thermodynamic averages, i.e. averages given by equilibrium measures). Similar tools has been derived for systems close to equilibrium (in local equilibrium) by means of linear irreversible thermodynamics and linear response theory. Although this task has been impossible to take further away from equilibrium, by using a similar approach we have been able to \emph{model} expressions representing some non-equilibrium situations in a non-local regime. The application of this formalism (based on the assumption of stochastic evolution at the mesoscopic level of description) permits a systematic study of such dissipative systems and in the cases outlined give rise to good agreement with available experimental data. Much work has to be done in order to fully understand the consequences of this and related formalisms, such as the ergodic dynamic systems approach of Gallavotti and Cohen \cite{galaco} and the (much debated) fractional calculus approach of Tsallis \cite{tsallis}. \emph{Acknowledgements}: One of the authors (E.H.L.) greatly appreciates the fruitful discussion with Rubén D. Zárate on modular automorphisms in arbitrary spaces.

\end{document}